\documentclass[aps,pra,reprint,raggedbottom,floatfix,superscriptaddress,longbibliography]{revtex4-2}
\usepackage[T1]{fontenc}
\usepackage{amsmath,amssymb,bm,graphicx,xcolor,booktabs}
\usepackage[colorlinks=true,allcolors=black]{hyperref}
\hypersetup{pdftitle={Correlations carried and controlled by coupled non-Markovian environments},pdfauthor={Stuart Flannigan}}
\newcommand{\dd}{\mathrm{d}}\newcommand{\E}{\mathbb{E}}\newcommand{\D}{\mathcal{D}}
\newcommand{\ket}[1]{\lvert #1\rangle}\newcommand{\bra}[1]{\langle #1\rvert}

\begin{document}
\title{Correlations carried and controlled by coupled non-Markovian environments}
\author{Stuart Flannigan}
\email{stuartflannigan@commutatorstudios.com}
\affiliation{Commutator Studios GmbH, 40211 D\"{u}sseldorf, Germany}
\date{November 10, 2026}
\begin{abstract}
An environment can retain quantum information and carry it between spatially separated systems. Describing the combined influence of these processes requires a treatment of both environmental memory and spatial correlations. Here, we formulate a local hierarchy of pure states for quantum systems coupled to spatially correlated, non-Markovian environments whose modes themselves exchange excitations. We use this method to describe the reduced spin dynamics of a chain of locally coupled Rabi spin--cavity pairs motivated by cavity-QED implementations, where photon hopping in the eliminated cavity environment generates correlations and entanglement between otherwise isolated spins. Monitoring intermediate cavities and feeding their measurement records back onto the corresponding spins further enhances correlations between endpoint spins that receive no direct feedback. The delayed response when feedback starts, and continued entanglement growth after it stops, support a mechanism in which coherent environmental dynamics carry the effects of earlier feedback to the endpoint spins. Finally, tensor-network simulations of XXZ and Heisenberg chains show how environmental transport changes many-body correlation patterns. These results show how coupled environments and local feedback can control correlations between distant spins. We provide a numerically exact stochastic pure-state framework for simulating these processes, opening up avenues to explore the coherent and dissipative control of interacting non-Markovian many-body systems.
\end{abstract}
\maketitle

\section{Introduction}
Non-Markovian environments remember earlier interactions with a quantum system and can return excitations after a delay~\cite{deVega2017}. When these excitations also propagate through the environment, they provide a route for correlations to develop between spatially separated parts of the system~\cite{Jing2015,Braun2002}. Controlling these environmental dynamics offers opportunities to influence correlation growth, but describing memory and transport together is particularly challenging for large, strongly interacting systems. Here we extend {the hierarchy of pure states (HOPS)~\cite{Suess2014}} to spatially coupled environments, retaining local system--environment interactions in a stochastic pure-state description. We use this method to investigate a chain of Rabi spin--cavity pairs, where photon hopping generates correlations and entanglement between spins with no direct interaction. Continuously monitoring intermediate cavities and feeding their output back onto the corresponding spins further enhances correlations between endpoints that are never actuated. The delayed response to starting and stopping feedback shows how these correlations build through the subsequent spin--cavity dynamics. Combining the hierarchy with tensor networks, we also show that environmental transport changes correlation spreading in interacting XXZ and Heisenberg chains. These results provide a way to study the control of non-Markovian mediated correlation growth through the coherent spatial dynamics, dissipative loss, and monitored output of an environment.

{
Coherent interactions~\cite{Kitagawa1993}, engineered dissipation~\cite{Kraus2008,Verstraete2009,Clark2003,Debecker2026StructuredBath}, and measurement-based feedback~\cite{Wiseman1993,Wang2005} offer complementary ways to control correlations and entanglement. In optical cavities, light-mediated interactions have generated collective spin squeezing~\cite{Leroux2010} and programmable spatial correlations~\cite{Periwal2021}. Engineered dissipation has produced entanglement between separated atomic ensembles~\cite{Krauter2011} and stabilised entangled states in superconducting circuits and trapped ions~\cite{Shankar2013,Lin2013}, while measurement and feedback have further enabled deterministic entanglement generation in circuit QED~\cite{Riste2013}. Continuous measurement and feedback have also stabilised the dynamics of a single superconducting qubit~\cite{Vijay2012}, while non-Markovian monitoring opens further possibilities when the measured field retains memory~\cite{Link2022}. Here, we examine how these ingredients work together when local environments communicate: the measured output of one cavity can carry information acquired elsewhere in the network.

Cavity QED provides a concrete setting for this question. Raman-assisted atom--cavity interactions can realise effective Rabi couplings~\cite{Grimsmo2013}, and Rabi-model quantum simulation and coherent coupling between resonators have been demonstrated in circuit QED~\cite{Braumuller2017,Raftery2014}. The cavity lifetime, spin--cavity coupling and photon hopping set independently adjustable timescales for information loss, local exchange and spatial propagation. Such networks can therefore probe the crossover from rapidly relaxing environments, where a memoryless spin description is appropriate, to environments whose dynamics must be retained. They are also useful quantum simulators of a broader problem: interacting matter coupled to local bosonic degrees of freedom. The Hubbard--Holstein model of electrons and lattice vibrations is one example; previous many-body HOPS calculations included damped local phonons~\cite{Flannigan2022}, but communicating environments add the possibility of propagation between those local modes. 
}

While generalised HEOM~\cite{Ikeda2020}, matrix-coupled hierarchies~\cite{Guo2025Extended,Xu2023QDMESS}, time-evolving matrix product operators~\cite{Strathearn2018} and pseudomode embeddings, including interacting modes~\cite{Garraway1997,Tamascelli2018,Pleasance2020,Muller2026}, allow for the numerical simulation of related descriptions of environmental memory and spatial correlations, HOPS describes non-Markovian dynamics using stochastic wavefunctions~\cite{Diosi1998,Suess2014,Hartmann2017} and has been shown to be suitable for a combination with tensor network methods~\cite{Flannigan2022}. However, typical local constructions assume independent memory modes, excluding direct transport between them on system timescales. Our construction retains coherent coupling between bath modes while keeping the system--bath interactions local. Our contribution is therefore a local HOPS formulation for communicating environments, including their monitored output and feedback, and a demonstration of its application to remote correlation control and interacting many-body dynamics. 

{
We first introduce the local spin--cavity model and present the correlated hierarchy in Sec.~\ref{sec:hierarchy}. Section~\ref{sec:transport} shows how photon hopping generates spin entanglement and correlations within the bath. In Sec.~\ref{sec_feed}, we examine how local measurement and feedback change remote spin correlations, including the effects of controller placement and switching. Section~\ref{sec:manybody} extends the calculation to interacting spin chains using tensor networks. We conclude in Sec.~\ref{sec:discussion}; derivations, numerical checks and additional mechanism controls are given in the Supplemental Material~\cite{SupplementalMaterial}.
}

\section{Local hierarchy for a communicating environment}\label{sec:hierarchy}
\subsection{Physical model}
Consider an open chain with one spin and one cavity at each site. With $\hbar=1$ and $S_i^\alpha=\sigma_i^\alpha/2$, the anisotropic Rabi Hamiltonian is
\begin{align}
 H_{\rm R}&=\omega_s\sum_i S_i^z+\sum_{ij}h_{ij}a_i^\dagger a_j\nonumber\\
 &\quad+\sum_i(F_i a_i^\dagger+F_i^\dagger a_i),\label{eq:rabi}\\
 h_{ij}&=\omega_c\delta_{ij}+J_b(\delta_{i,j+1}+\delta_{i,j-1}).\label{eq:bath}
\end{align}
Here $F_i=g_-S_i^-+g_+S_i^+$. There is no direct spin--spin interaction in this model: spin $i$ couples only to cavity $i$, and tunnelling occurs only between cavity modes. The Jaynes--Cummings and full Rabi limits are $g_+=0$ and $g_+=g_-$, respectively. Each cavity loses photons into an independent vacuum reservoir, giving the GKSL master equation~\cite{Gorini1976,Lindblad1976}
\begin{equation}
 \begin{aligned}
 \dot\varrho&=-i[H_{\rm R},\varrho]+\sum_i\kappa_i\D[a_i]\varrho,\\
 \D[A]\varrho&=A\varrho A^\dagger-\tfrac12\{A^\dagger A,\varrho\}.
 \end{aligned}
 \label{eq:loss}
\end{equation}
Here $\kappa_i$ is the photon-number decay rate and $\gamma_i=\kappa_i/2$ the amplitude decay rate. Throughout, the spins constitute the system and the coupled cavities its environment. Including the cavities in an enlarged system gives the Markovian master equation above; eliminating them leaves spin dynamics that retain the cavities' memory. The existence of this enlarged description does not mean that the spin dynamics themselves are Markovian. More generally, in the field of open quantum systems, enlarging the ``system'' to retain some of the environmental degrees of freedom often gives a Markovian description even when its reduced dynamics are non-Markovian~\cite{deVega2017,Pollock2018}. Below, we treat an engineered effective Rabi Hamiltonian and an interaction quench from an initially uncorrelated spin state and cavity vacuum, simulating the reduced dynamics of the spin system, but retaining bath evolution and system--bath correlations without a Born--Markov approximation for the spin--cavity coupling.

An optical implementation can encode each spin in two long-lived atomic states, coupled through far-detuned Raman transitions and a cavity mode~\cite{Grimsmo2013}. Laser amplitudes and detunings set the effective $g_\pm$, $\omega_s$ and $\omega_c$; these frequencies need not be the bare optical transition frequencies. Eliminating the excited atomic levels requires detunings large compared with their couplings and decay rates. With Markovian output reservoirs, cavity leakage is then described by Equ.~\ref{eq:loss}, and coupled resonators provide the additional multi-spin--cavity design for our considered configuration~\cite{Raftery2014}.

The vacuum photon mode correlation is fixed by the drift matrix $W_{ij}=\gamma_i\delta_{ij}+ih_{ij}$:
\begin{equation}
 K_{ij}(\tau)=\langle a_i(\tau)a_j^\dagger(0)\rangle=[e^{-W\tau}]_{ij},\qquad\tau\geq0.
 \label{eq:kernel}
\end{equation}
Here $e^{-W\tau}$ is a matrix exponential. Initially $K_{ij}(0)=\delta_{ij}$. Hopping of photons between cavities produces delayed offsite correlations and changes onsite return amplitudes. Thus changing $J_b$ changes both spatial transport and local memory; it is not generally a comparison with matched local spectra. If the cavity correlations decay much faster than the spin response, eliminating them can give an approximately Markovian spin equation~\cite{Damanet2019}. When these timescales are comparable, the returning and propagating field must instead be accounted for. Our hierarchy retains this response without making that separation-of-timescales approximation.

\subsection{Correlated HOPS}
{
Tracing out a Gaussian environment does not remove its influence from the dynamics: it appears as correlated fluctuations and a response to the system's past evolution. Non-Markovian quantum state diffusion (QSD) expresses this exactly in terms of stochastic system wavefunctions~\cite{Diosi1998}. For the initial vacuum and factorised preparation used here, its linear form is
\begin{equation}
 \partial_t\ket{\psi_t}=\left[-iH_S+\sum_iF_i z_i^*(t)-\sum_iF_i^\dagger D_i(t)\right]\ket{\psi_t}.
 \label{eq:qsd_main}
\end{equation}
The complex Gaussian noise has zero mean, covariance $\E[z_i(t)z_j^*(u)]=K_{ij}(t-u)$ for $t\geq u$, where $K_{ij}(\tau)$ is defined in Equ.~\ref{eq:kernel}. The indices $i,j$ correspond to different bath modes coupled to the system through the system operators $F_i,F_j$ defined to be local in our case considered here. The memory operator acts on the causal wavefunction as
\begin{equation}
 D_i(t)\ket{\psi_t}=\sum_j\int_0^t K_{ij}(t-u)
 \frac{\delta\ket{\psi_t}}{\delta z_j^*(u)}\,\dd u.
 \label{eq:memory_main}
\end{equation}
The functional derivative measures the response to an earlier noise value. Its weighted integral therefore carries the bath's response to previous system dynamics, including propagation from bath mode $j$ to bath mode $i$, and therefore, modifications to the bath caused by system site $j$ can propagate through the bath modes and influence the dynamics of system site $i$.

HOPS replaces the functional memory term by auxiliary wavefunctions~\cite{Suess2014}. Starting with the root $\ket{\psi_{\bm0}}=\ket{\psi_t}$, define the factorially scaled tiers
\begin{equation}
 \ket{\psi_{\bm k}}=\frac{\prod_i D_i^{k_i}}{\sqrt{\prod_i k_i!}}\ket{\psi_t}.
 \label{eq:tiers_main}
\end{equation}
The first tier stores one bath response, and higher tiers store repeated responses. Differentiating the tiers produces decay and mixing terms governed by $W$. For example, the product rule for the first tier, $\ket{\psi_{\bm e_i}}=D_i\ket{\psi_{\bm0}}$, gives
\begin{align}
 \partial_t\ket{\psi_{\bm e_i}}
 &=(\partial_tD_i)\ket{\psi_{\bm0}}+D_i\partial_t\ket{\psi_{\bm0}}\nonumber\\
 &=-\sum_jW_{ij}\ket{\psi_{\bm e_j}}+D_i\partial_t\ket{\psi_{\bm0}}.
 \label{eq:first_tier_product_main}
\end{align}
Here $\bm e_i$ is the unit multi-index. The second line uses $\partial_tD_i=-\sum_jW_{ij}D_j$, which follows from $\partial_tK=-WK$ using the full-noise-history operator convention described in Supplement Sec.~I\,A. The first term therefore describes decay and transfer of the stored bath response. Substituting Equ.~\ref{eq:qsd_main} into the remaining derivative couples this first tier to the root and second tiers. Repeating the same product rule generates the higher tiers. We collect these wavefunctions in an auxiliary basis,
\begin{equation}
 \ket{\Psi}=\sum_{\bm k}\ket{\bm k}\ket{\psi_{\bm k}},\qquad
 b_i\ket{\bm k}=\sqrt{k_i}\ket{\bm k-\bm e_i}.
 \label{eq:auxiliary_basis_main}
\end{equation}
The resulting linear hierarchy is local in time,
\begin{align}
 \partial_t\ket{\Psi}&=\mathcal G\ket{\Psi},\nonumber\\
 \mathcal G&=-iH_S+\sum_iF_i z_i^*\nonumber\\
 &\quad+\sum_i(F_i b_i^\dagger-F_i^\dagger b_i)
 -\sum_{ij}W_{ij}b_i^\dagger b_j.
 \label{eq:linear_hierarchy_main}
\end{align}
Each tier is a spin wavefunction, and its indices are memory coordinates, not a physical photon-number basis. Initially only the root tier $\bm k=\bm0$ is populated, and these higher auxiliary states become populated through the time evolution under the above differential equations, reflecting the deviation of the bath away from its initial state. Note that while the hierarchy is exact, in practice the hierarchy depth must be truncated to some $k_{\mathrm{max}}$, which becomes a numerical control parameter. For numerical sampling we also use the nonlinear form of QSD and HOPS~\cite{Diosi1998,Suess2014,Hartmann2017}, which shifts the noise according to the evolving root state and changes the sampling measure (Supplement Sec.~I\,B).

In Equ.~\ref{eq:linear_hierarchy_main}, each spin couples to its own hierarchy coordinate, while the off-diagonal entries of $W$ transfer the bath response between coordinates alongside the spatial correlations in the noise. This extends the many-body construction of Ref.~\cite{Flannigan2022} to the matrix kernel in Equ.~\ref{eq:kernel}.

Note that if the local bath properties are uniform, then a basis change to independent bath normal modes provides an alternative route using conventional HOPS~\cite{Varvelo2021,Citty2024,Lynd2025,Muller2025}. However, the distinction is where the spatial coupling appears: we keep communicating bath modes and \textit{local} spin--bath interactions, whereas the normal-mode representation has independent modes that generally couple to several spins. Keeping the physical local coordinates allows us more flexibility to vary the loss profile and local feedback without the correlated damping or spatially extended couplings that can arise in a normal-mode basis, and also allows us to retain the short-range structure used by the tensor-network calculation, following the same principle of locality that motivates Wannier-basis constructions of Hubbard models~\cite{Walters2013}.

With $\ket{\widetilde\psi_{\bm0}}$ denoting a normalised nonlinear root and $\E_{\rm nl}$ the average over the transformed sampling measure, the physical spin density is
\begin{equation}
 \rho_S(t)=\E_{\rm nl}\big[\ket{\widetilde\psi_{\bm0}(t)}\bra{\widetilde\psi_{\bm0}(t)}\big].
 \label{eq:root_density}
\end{equation}
The higher tiers only store the response needed to track memory and influence the system state; they are not physical cavity states to be traced out.
}

\begin{figure}[!t]
\includegraphics[width=\columnwidth]{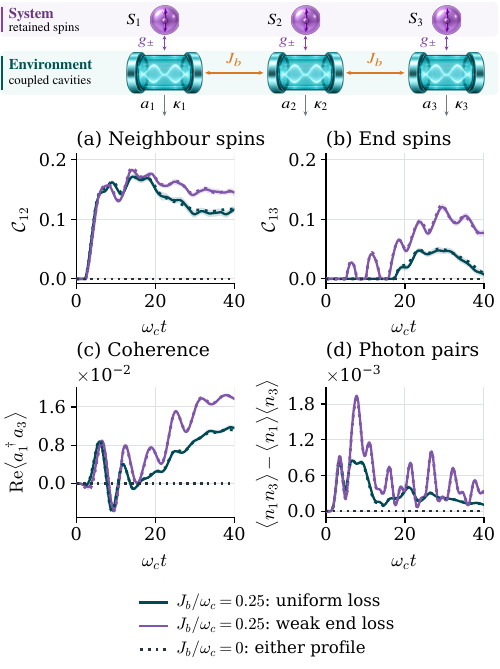}
\caption{\textbf{Local coupling, remote correlations.} Each spin couples to its own cavity, with photon hopping between cavities and no direct spin interaction. {The numerical plots show neighbouring concurrence in (a), endpoint concurrence in (b), real endpoint photon coherence in (c), and connected photon-number covariance in (d), with $n_i=a_i^\dagger a_i$.} Uniform number losses $(0.2,0.2,0.2)\omega_c$ and weak ends $(0.04,0.52,0.04)\omega_c$ have the same total loss. The hopping curves use $J_b=0.25\omega_c$; the $J_b=0$ cross correlations vanish by factorisation. Spins start in $\ket{ggg}$ and cavities in vacuum, with $\omega_s=0.1\omega_c$, $g_+/g_-=1/4$ and $g_-^2+g_+^2=0.18\,\omega_c^2$. Solid curves and pointwise 95\% bands are nonlinear HOPS; same-colour dotted curves are photon-cutoff-checked master equations. There is no measurement or feedback. Schematic artwork generated with AI.}
\label{fig:rabi}
\end{figure}

\section{Correlations carried by the bath}\label{sec:transport}
We begin with three spins in $\ket{ggg}$ and empty cavities. Counter-rotating terms create spin--photon excitations after the interaction quench and the photon hopping allows the subsequent dynamics to involve another spin, even though in Equ.~\ref{eq:rabi} spins are uncoupled. Fig.~\ref{fig:rabi} compares uniform losses (teal) with weaker loss at the ends (purple), keeping the total loss fixed. Without hopping the local pairs remain independent, as shown by the black dotted controls. With hopping, both neighbouring spins [{Fig.~\ref{fig:rabi}(a)}] and endpoint spins [{Fig.~\ref{fig:rabi}(b)}] become entangled. We quantify this using the two-spin concurrence~\cite{Wootters1998},
\begin{equation}
 \mathcal C_{ij}=\max(0,\lambda_1-\lambda_2-\lambda_3-\lambda_4),
 \label{eq:concurrence}
\end{equation}
where the $\lambda_a$ are the decreasing square roots of the eigenvalues of $\rho_{ij}\widetilde\rho_{ij}$, with $\widetilde\rho_{ij}=(\sigma^y\otimes\sigma^y)\rho_{ij}^*(\sigma^y\otimes\sigma^y)$. It is zero for separable two-spin states and one for a maximally entangled pair. The endpoints communicate through two photon links, although neither spin couples directly to its neighbour's cavity.

The larger endpoint concurrence for weak-end loss in Fig.~\ref{fig:rabi}(b) (purple) reflects the longer memory available at the edges, allowing more bath excitations to be returned to those endpoint spins. Additionally, the reduction in endpoint loss prolongs the bare antisymmetric bath mode $(a_1-a_3)/\sqrt{2}$, which has no middle-cavity component. Both mechanisms give rise to the observed concurrence enhancement, see the normal-mode analysis in Supplement Sec.~IV for more details.

The photon observables in Fig.~\ref{fig:rabi}(c,d) give a complementary view of this communication. Hopping builds end-cavity coherence and connected number fluctuations, both have a larger response for weak-end loss (purple) than for uniform loss (teal). These describe the occupied, interacting environment and are distinct from the free vacuum kernel in {Equ.~\ref{eq:kernel}}. Populations can grow without hopping, whereas these cross correlations vanish. Their development alongside spin entanglement connects the spin response to correlations carried by the environment.

The weak-end comparison also tests the hierarchy when a bath basis change cannot separate the full response into independent damped modes. This is the parameter regime in which retaining the coupled bath response exceeds the capabilities of conventional independent-mode HOPS constructions~\cite{Suess2014,Hartmann2017,Muller2025}; see Supplement Fig.~S4 for an explicit comparison.

\section{Local feedback controls remote correlations}\label{sec_feed}
\subsection{Continuous monitoring and local actuation}
Cavity leakage is both a loss channel and a source of information. The outgoing field has interacted with its local spin and, when $J_b\ne0$, with neighbouring cavities. Measuring it while the system evolves therefore gives a signal with which to control the subsequent dynamics. We use continuous homodyne detection and local feedback, following cavity-QED control schemes~\cite{Wiseman1993,Wang2005,Link2022}.

In homodyne detection, the leaking field is mixed with a strong coherent local oscillator, and a classical electronic controller uses this to modulate a coherent drive on the spin, for example a laser or Raman drive in an atomic implementation, or a microwave drive in circuit QED. Here the drive amplitude is proportional to the instantaneous current, producing rotations generated by $S_j^x$. The loop returns a classical signal to the spin, not the emitted photons. Continuous measurement and electronic feedback of this kind have experimental precedents in circuit QED~\cite{Vijay2012}.

For each monitored site $j$, the output operator $c_j$, local feedback operator $G_j$, homodyne record $Y_j$ and feedback rotation $U_j$ are
\begin{align}
 c_j&=e^{-i\phi}\sqrt{\kappa_j}a_j,\qquad G_j=f_jS_j^x,\nonumber\\
 \dd Y_j&=\sqrt{\eta_j}\langle c_j+c_j^\dagger\rangle_Y\dd t+\dd W_j,
 \qquad U_j=e^{-iG_j\dd Y_j}.\label{eq:record}
\end{align}
Here $\eta_j\in[0,1]$ is the detection efficiency, $\phi$ selects the measured field quadrature, and $\langle\cdot\rangle_Y$ denotes an expectation conditioned on the measurement record. The independent real Wiener increments obey $\E[\dd W_j\dd W_k]=\delta_{jk}\dd t$. The gain $f_j$ has units $\sqrt{\omega_c}$, so $G_j\dd Y_j$ is dimensionless. Equ.~\ref{eq:record} gives the small-time spin rotation applied after each measurement increment. We use ideal zero-delay feedback, with $\phi=\pi/2$ and $\eta_j=1$ unless stated otherwise. Experimentally, this requires the detector and controller response to be fast compared with the spin--cavity dynamics; finite bandwidth and delay are not included here. 

Averaging this measurement-and-feedback sequence over the physical records gives the standard additional contribution to the spin--cavity master equation~\cite{Wiseman1993},
\begin{equation}
 \mathcal L_{\rm fb}\varrho=\sum_j\left\{\D[G_j]\varrho-i\sqrt{\eta_j}[G_j,c_j\varrho+\varrho c_j^\dagger]\right\}.
 \label{eq:feedback}
\end{equation}
Here $\D$ is the Lindblad dissipator in Equ.~\ref{eq:loss}, the sum runs over monitored and actuated sites, and $\varrho$ is the joint state averaged over records. The original cavity-loss terms remain present.

The dissipator $\D[G_j]$ describes the effect of the white-noise part of the drive, while the commutator in Equ.~\ref{eq:feedback} retains the correlation between the measured output and actuation. Driving the same operators with independent white noise retains only $\D[G_j]$. Monitoring without feedback, when the record is discarded, leaves the unconditional state unchanged. 

To simulate this equation after eliminating the cavities, the hierarchy must retain the correlation between the cavity field and its outgoing signal. For ideal detection at all loss ports, with zero gain on unactuated ports, define $\ell_{rj}=e^{-i\phi_r}\sqrt{\kappa_r}\delta_{rj}$ and sample the joint complex Gaussian process
\begin{align}
 \dd z&=-Wz\dd t-\ell^\dagger\dd B, &
 \dd\Xi&=\dd B+\ell z\dd t.\label{eq:joint_output_main}
\end{align}
Here $B$ is a vector of independent proper complex Wiener processes and $z(0)$ is a stationary unit-covariance complex Gaussian vector. The output increments are white, but have a retarded correlation with the earlier cavity field. Together with the corresponding cross term in the hierarchy, this reproduces {Equ.~\ref{eq:feedback}}. See Supplement Sec.~II\,C for more details. 

\begin{figure}[!t]
\includegraphics[width=\columnwidth]{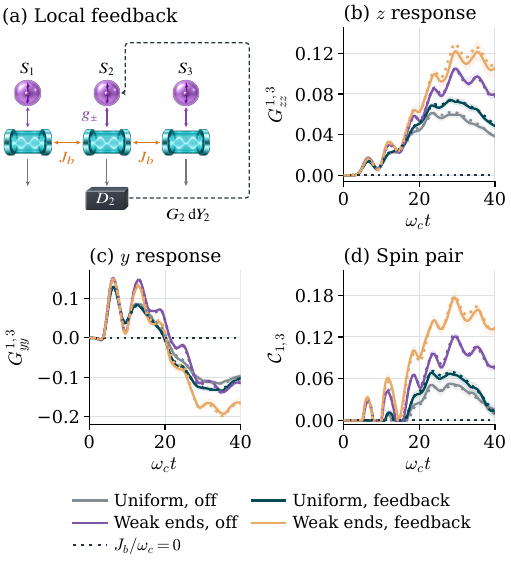}
\caption{\textbf{An intermediate controller enhances endpoint correlations.} {(a):} Homodyne detection of cavity 2 supplies a classical signal that drives only spin 2. {(b,c):} Signed connected endpoint $zz$ and $yy$ correlations. {(d):} Concurrence of the averaged endpoint state. Uniform/weak-end damping and feedback off/on are compared with the zero-hopping control. Parameters match Fig.~\ref{fig:rabi}; the feedback gain is $f_2=0.15\sqrt{\omega_c}$, phase $\phi=\pi/2$ and detection efficiency $\eta_2=1$. Solid nonlinear HOPS curves have pointwise 95\% bands; same-colour dotted lines are matched finite-cutoff master equations. Neither spin endpoint is actuated. Schematic artwork generated with AI.}
\label{fig:feedback3}
\end{figure}

The physical measurement acts on photons leaking from the cavities into their Markovian output reservoirs. The spins nevertheless interact with a non-Markovian cavity environment. The real $Y_j$ in Equ.~\ref{eq:record} is the experimentally measured current integrated in time. The complex noises $z$ and $\Xi$ instead provide a computational sampling of the averaged feedback dynamics: our output-HOPS paths sample the unconditional reduced spin density and are not individual homodyne measurement records. Alternatively, a hierarchy conditioned on physical measurement records is developed in Ref.~\cite{Link2022}.

Feedback-enhanced entanglement and non-Markovian monitoring have important precedents~\cite{Wang2005,Martin2015,Yamamoto2008,Link2022}. Link \emph{et al.}\ derive conditioned hierarchies for multimode cavities, study measurement-conditioned correlations between separated atomic groups and demonstrate feedback-enhanced collective spin squeezing. Here the target is the unconditional endpoint state under local actuation of an intermediate spin. The question is how a network of local, communicating environments mediates control of spins that are not themselves actuated.

\subsection{Feedback enhancement across the bath network}

\begin{figure}[!t]
\includegraphics[width=\columnwidth]{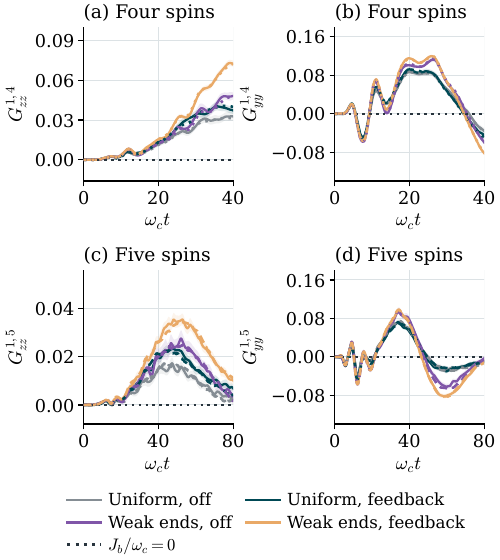}
\caption{\textbf{Correlations persist across a longer bath network.} {Endpoint $zz$ and $yy$ correlations for four (a,b) and five (c,d) spins}, retaining uniform/weak-end losses and feedback off/on. {(c)} uses a tighter vertical scale to resolve the smaller five-spin signal. Weak-end number losses are $(0.04,0.36,0.36,0.04)\omega_c$ for four sites and $(0.04,0.92/3,0.92/3,0.92/3,0.04)\omega_c$ for five; uniform losses are $0.2\omega_c$ at each site. Site 2 is controlled for $N=4$, and site 3 for $N=5$, with the gain and phase of Fig.~\ref{fig:feedback3}. Solid curves and pointwise 95\% bands are HOPS. Dotted four-spin references are finite-cavity master equations. Neither spin endpoint is actuated.}
\label{fig:reach}
\end{figure}

We characterise the remote response in Figs.~\ref{fig:feedback3} and \ref{fig:reach} through connected correlations of the unconditional spin state,
\begin{equation}
 G_{ij}^{\alpha\beta}=\langle\sigma_i^\alpha\sigma_j^\beta\rangle-
 \langle\sigma_i^\alpha\rangle\langle\sigma_j^\beta\rangle.\label{eq:pair_connected}
\end{equation}
Both the pair moment and the single-spin means are accessible through repeated local readout with appropriate basis rotations. In the simulation they are formed after averaging the root density, rather than by averaging the connected correlation of each trajectory. Concurrence is likewise evaluated on the averaged two-spin state.

The local controller in Fig.~\ref{fig:feedback3}(a) enhances correlations between the two endpoints. For weak-end loss, feedback (orange) increases the $zz$ response in {Fig.~\ref{fig:feedback3}(b)} and the concurrence in {Fig.~\ref{fig:feedback3}(d)} compared with feedback off (purple). Under uniform loss, the corresponding teal and grey curves show a smaller enhancement. The $yy$ curves in {Fig.~\ref{fig:feedback3}(c)} also change sign during the evolution, with feedback producing a larger late-time magnitude. These different responses show how local control reshapes the correlations established by the bath. 

Without photon hopping, the separated spins remain uncorrelated even with feedback: the black dotted curves in Fig.~\ref{fig:feedback3}(b--d) stay at zero. The enhancement therefore depends on the combined action of transport between cavities, local spin--cavity exchange and feedback conditioned on the measured output.
{
The weak-end response in Fig.~\ref{fig:feedback3} combines two roles of the loss profile. Weak endpoint damping prolongs the memory available for spin--cavity exchange, while the more strongly damped middle cavity couples its field more strongly to the monitored output. {Supplement Sec.~III\,C examines these roles} by comparing the weak-end network with uniform loss reduced to the same endpoint value. Although uniformly low loss gives a larger concurrence peak, feedback in the weak-end network sustains greater endpoint entanglement over a later interval [{Fig.~S2(c)}, orange and purple curves]. These controls suggest that stronger access to the central output helps feedback maintain the correlations carried by the coupled cavities, while weak endpoint loss slows their decay.
}

Feedback also enhances connected correlations across a greater separation between the endpoints. Fig.~\ref{fig:reach}(a,b) shows the four-spin chain and {Fig.~\ref{fig:reach}(c,d)} the five-spin chain, each with one intermediate controller and local spin--cavity coupling. The mean loss is kept fixed: the endpoints remain weakly damped and the remaining loss is shared among the interior cavities. {For the $zz$ correlations in Fig.~\ref{fig:reach}(a,c)}, feedback in the weak-end network (orange) strengthens the response compared with feedback off (purple). The smaller five-spin signal shows the reduced effect across a longer bath network, while {the $yy$ correlations in Fig.~\ref{fig:reach}(b,d)} retain a pronounced dynamical response. This indicates that these bath-induced system correlations persist as the separation between spins grows, although the magnitudes of the correlations are of course reduced.

\begin{figure}[!t]
\includegraphics[width=\columnwidth]{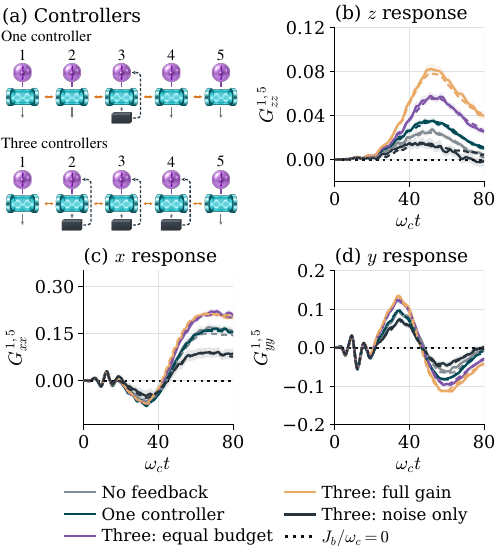}
\caption{\textbf{Distributed local feedback reshapes remote correlations.} {(a):} One or three interior detector--actuator pairs in the five-spin weak-end chain. {(b--d):} Endpoint $zz$, $xx$ and $yy$ correlations. One controller acts on site 3; three controllers act independently on sites 2, 3 and 4. Full gain uses $f_j=0.15\sqrt{\omega_c}$ at every controlled site; the equal-budget label denotes equal summed squared gain, with $f_j=0.15\sqrt{\omega_c/3}$. Independent full-gain drive noise and no-feedback curves distinguish the measured-record contribution. All physical parameters otherwise match Fig.~\ref{fig:reach}. Solid curves and bands are HOPS with pointwise 95\% uncertainty. Dashed curves are independent finite-cutoff cavity-SSE means, with separate sampling uncertainty. Neither spin endpoint is actuated. Schematic artwork generated with AI.}
\label{fig:controllers}
\end{figure}

\subsection{Several controllers}

The additional spin--cavity sites also provide new places to monitor and actuate the chain. This makes the longer network a setting in which to ask whether distributing the control can compensate for the smaller remote response. In Fig.~\ref{fig:controllers}(a), all three interior spins receive feedback from their own cavity outputs. We compare full gain on each controlled site with a choice that fixes $\sum_j f_j^2$ to its single-controller value, labelled equal budget in the figure. This matches the summed squared gain; the three-controller case also uses three monitored outputs. The full-gain curves (orange) show a larger intermediate-time $zz$ response in {Fig.~\ref{fig:controllers}(b)} than one controller (teal), while the equal-budget response (purple) lies between them. {Fig.~\ref{fig:controllers}(c,d)} show that the controller arrangement also changes the $xx$ and $yy$ components. Thus the choice of measured outputs and local feedback operators gives control over both the strength and the form of the remote correlations.

Driving the same local spin operators with independent white noise at the same gains (charcoal), instead of the measured homodyne records, does not reproduce the feedback response (orange) in Fig.~\ref{fig:controllers}(b--d). Therefore, the relation between each output field and its local actuator is central to the enhancement. Together, these comparisons show how the selected outputs and local gains reshape bath-mediated correlations between distant, noninteracting endpoint spins. {Supplement Fig.~S3} further compares detection efficiency and controller signs, showing that how the available output is used also matters.

\begin{figure}[tbp]
\includegraphics[width=\columnwidth]{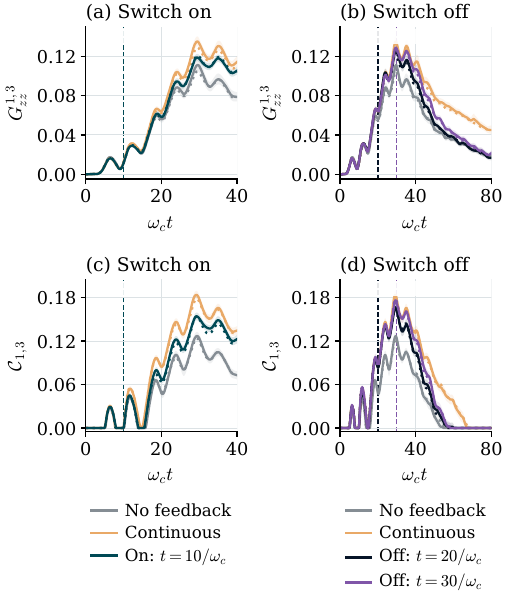}
\caption{\textbf{The endpoint response retains the history of feedback.} {Three-spin weak-end chain. Feedback on spin 2 is activated at $t=10/\omega_c$ (a,c) or stopped at $t=20/\omega_c$ or $30/\omega_c$ (b,d), compared with never/always controls.} Colour-matched vertical dashed guides mark the changes; each column's legend applies to both panels in that column. The upper row shows signed $G_{13}^{zz}$ and the lower row the concurrence of the averaged endpoint state. The full joint state is retained through each switch. Solid curves and pointwise 95\% bands are checked HOPS estimates; same-colour dotted curves are master equations. The delayed-on record ends at $t=40/\omega_c$, and the stop comparisons extend to $t=80/\omega_c$. Preparation, loss profile and controller match Fig.~\ref{fig:feedback3}.}
\label{fig:mechanism}
\end{figure}

\subsection{Feedback history and the delayed response}
We now analyse the effects of starting and stopping the feedback protocol. In Fig.~\ref{fig:mechanism} we use this process to show how local feedback changes remote correlations through the subsequent dynamics. In {Fig.~\ref{fig:mechanism}(a,c)}, delayed activation (teal) initially follows the no-feedback curve (grey), then moves towards the continuous-feedback response (orange) after the dashed switch-on line. Conversely, stopping feedback at $\omega_ct=20$ (blue) or $\omega_ct=30$ (purple) in {Fig.~\ref{fig:mechanism}(b,d)} does not immediately remove its effect. The endpoint concurrence in {Fig.~\ref{fig:mechanism}(d)} can continue to rise after the switch before decaying. The instantaneous setting of the controller therefore does not determine the instantaneous correlation growth.

A local, trace-preserving operation on the middle spin cannot instantaneously change the reduced endpoint state. Feedback instead modifies the joint spin--cavity state. The measured central field has already interacted with neighbouring cavities, so a rotation conditioned on that field can change correlations that involve more than the local pair. The mixed spin--cavity correlations and source calculation {in Supplement Sec.~III\,B and Fig.~S1} support this picture. Subsequent spin--cavity exchange and photon hopping carry these changes back to the endpoints. Stopping feedback leaves the previously prepared joint state intact, allowing its later evolution to continue building endpoint correlations. After feedback stops, the dynamics trend back towards the no-feedback case over the displayed interval, see Fig.~\ref{fig:mechanism}(b,d). The switch-on and switch-off results thus support a dynamical accumulation of the effects of earlier measurement and feedback. Spin--cavity entanglement alone does not explain the remote enhancement: the measured field is part of a communicating environment. See Supplement Sec.~III for more details.

{
These switching results add a temporal test of the loss-profile mechanism discussed above. The central output can influence the later endpoint response even after actuation has stopped, whereas turning feedback on produces no immediate increase at the endpoints. Thus the persistence of correlations cannot be attributed solely to the current feedback strength: the spin--cavity state prepared by earlier monitoring and feedback also matters. Photon hopping and local spin--cavity exchange then carry this influence through the network.
}

\section{Many-body dynamics}\label{sec:manybody}
We now add direct nearest-neighbour spin interactions to the same local spin--cavity geometry. Correlations can therefore propagate both through the spin chain and through the coupled environment. We replace the few-spin Zeeman term by
\begin{equation}
 H_S=J\sum_{i=1}^{N-1}(S_i^xS_{i+1}^x+S_i^yS_{i+1}^y+\Delta S_i^zS_{i+1}^z).
 \label{eq:xxz}
\end{equation}
We consider a transverse coupling $F_i=gS_i^x$ to an XXZ chain and local excitation exchange $F_i=gS_i^-$ in the Heisenberg limit. Both begin in a $z$-N\'eel state. For twenty spins these preparations probe finite-density dynamics beyond a one-excitation description. An interleaved spin--auxiliary matrix-product state represents the nonlinear HOPS hierarchy~\cite{Flannigan2022,Gao2022,Fishman2022}. These calculations extend the unmonitored method; a many-body feedback calculation is a separate step left for a future study.

\begin{figure}[tbp]
\includegraphics[width=\columnwidth]{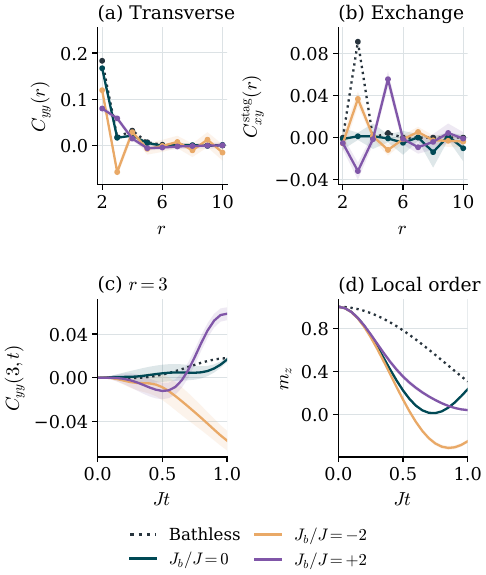}
\caption{\textbf{Environmental transport changes many-body correlations.} Twenty-spin nonlinear HOPS--MPS calculations with a $z$-N\'eel initial state. The transverse XXZ model has $\Delta=2$, $F_i=gS_i^x$ and $\omega_c=5J$; the exchange Heisenberg model has $\Delta=1$, $F_i=gS_i^-$ and zero rotating-frame cavity detuning. Both use $g=2J$, amplitude loss $\gamma=0.1J$ and $J_b/J=0,\pm2$. {(a,b):} Signed spatial correlation profiles at $Jt=1$. {(c):} Transverse $yy$ correlation at separation three. {(d):} Staggered order for exchange coupling. Every panel retains its bathless control. Bands are pointwise 95\% normal-approximation intervals.}
\label{fig:manybody}
\end{figure}

For diagonal spin components, we average $G_{i,i+r}^{\alpha\alpha}$ over spatial origins. The mixed exchange correlation also includes the N\'eel sublattice sign,
\begin{equation}
 C_{xy}^{\rm stag}(r,t)=\frac{1}{N-r}\sum_{i=1}^{N-r}(-1)^{i-1}G_{i,i+r}^{xy}(t).
 \label{eq:connected}
\end{equation}
{
{Here $G_{ij}^{xy}=\langle\sigma_i^x\sigma_j^y\rangle-\langle\sigma_i^x\rangle\langle\sigma_j^y\rangle$ is the connected correlation between the $x$ component of spin $i$ and the $y$ component of spin $j$. The factor $(-1)^{i-1}$ accounts for the alternating N\'eel sublattices. Local order is measured by the staggered magnetisation $m_z(t)=N^{-1}\sum_i(-1)^{i-1}\langle\sigma_i^z(t)\rangle$.}
}

Fig.~\ref{fig:manybody} shows that bath hopping changes the spatial pattern of spin correlations as well as the decay of local order. The transverse and exchange models respond differently, and positive and negative hopping can select different signs of the mixed correlation. The environment therefore supplies more than a uniform amplification or damping of the isolated-chain response. This extends the role of non-Markovian memory in correlation spreading~\cite{Flannigan2022} to a bath that also transports excitations between sites.

The hopping-sign comparison is especially useful for separating memory from spatial propagation. On a bipartite bath chain, $J_b\to-J_b$ leaves the onsite free-bath kernels unchanged and multiplies each offsite kernel by $(-1)^{i-j}$. With spin exchange and local spin--bath couplings held fixed, this sign changes the relative phase around a spin--bath plaquette. The contrast is therefore consistent with interference between direct spin transport and propagation through the environment. The selected short-time profiles demonstrate this change in correlation pattern; representative numerical refinements support those observables, while more distant tails remain sampling limited.

\section{Discussion}\label{sec:discussion}
An environment coupled locally to each spin can nevertheless provide a route for remote communication, here mediated through photon hopping which generates correlations and entanglement between spins with no direct interaction. The ability to simulate this without explicitly retaining the bath basis states requires a method that extends well beyond standard Markovian approximations: the photon environment not only retains memory of previous interactions with the system, but can transport that excitation/memory to spatially separated parts of that system at later times. Our hierarchy describes this problem with spin wavefunctions and auxiliary memory states, without a Born--Markov approximation for the spin--cavity coupling. 

Measuring an intermediate output and feeding it back onto its local spin changes this communication and can further enhance endpoint correlations. The delayed response when feedback starts, together with continued entanglement growth after it stops, shows why the history of the joint state matters: previous control changes the state on which later bath-mediated dynamics act. Additionally, non-uniform loss provides a way to control this mechanism: weak endpoint damping retains local memory, while stronger central damping couples the correlated field to the detector. Our loss and efficiency controls support the interpretation that balancing these roles can sustain greater feedback-enhanced correlations over a later interval. This makes spatially resolved dissipation a useful control parameter alongside hopping, detector placement and feedback gain.

The numerical contribution is a local HOPS representation that treats correlated fluctuations and bath response together. Keeping local spin--bath couplings while allowing unequal cavity losses makes it possible to explore these control profiles directly, including cases that cannot be reduced to independent damped normal modes. The tensor-network examples show how the same local organisation can be efficiently extended to interacting many-body systems, where strong non-Markovian environmental effects can lead to transport which competes with direct system dynamics.

{
The cavity network is thus both an experimentally motivated setting for remote correlation control and a model of communicating bosonic environments. It suggests tuning cavity loss, hopping and local homodyne feedback to study when environmental memory supports correlation growth and when leakage removes it. The same local organisation also connects to condensed-matter models. Replacing the spin chain by a Hubbard lattice and coupling each electronic density to a local oscillator gives a Hubbard--Holstein-type setting. Previous HOPS and quantum-trajectory calculations retained the coherent oscillation and damping of independent local phonons~\cite{Flannigan2022,Moroder2023}; coupling those modes would additionally allow vibrational disturbances to propagate and influence electronic dynamics elsewhere~\cite{Jansen2022}. We do not simulate that fermionic problem here, but the present method provides a route to studying how environmental transport competes with transport within interacting matter. Extensions to structured continua and finite-bandwidth feedback would broaden this connection beyond the effective cavity model.
}

\par\vspace{0.8\baselineskip}
\section*{Data and code availability}
The numerical data underlying the figures and companion Python and Julia implementations of correlated HOPS are publicly available at \url{https://github.com/SFlann/correlated-hops}. Additional simulation and analysis code is available from the author upon reasonable request.

\begin{acknowledgments}
I thank C\'esar Rodr\'iguez-Rosario for reading the manuscript and for valuable comments that helped strengthen its motivation and presentation.

OpenAI Codex (Astra) assisted with checking mathematical derivations, developing and debugging simulation and analysis code, and coordinating simulation workflows, including job scheduling, monitoring and data organisation. It also assisted with manuscript preparation. The author directed and reviewed the work; numerical results were assessed through independent reference calculations and convergence tests. The author takes responsibility for the final content.
\end{acknowledgments}

\nocite{Guhne2009,Popp2005,Zhang2018}
\let\originalbibfont\bibfont\renewcommand{\bibfont}{\originalbibfont\setlength{\baselineskip}{0.98\baselineskip}}\bibliography{references}
\end{document}


\title{Supplemental Material: Correlations carried and controlled by coupled non-Markovian environments}
\author{Stuart Flannigan}
\email{stuartflannigan@commutatorstudios.com}
\affiliation{Commutator Studios GmbH, 40211 D\"{u}sseldorf, Germany}
\date{November 10, 2026}
\maketitle
\tableofcontents

\section{Correlated hierarchy, noise generation and physical observables}\label{sec:hierarchy}

\subsection{Functional derivatives and factorial scaling}
For a factorised pure preparation and a vacuum Gaussian environment, use the linear non-Markovian stochastic Schr\"odinger equation
\begin{equation}
 \partial_t\ket{\psi_t}=\left[-iH_S+\sum_iF_i z_i^*(t)-\sum_iF_i^\dagger D_i(t)\right]\ket{\psi_t},\qquad
 D_i(t)=\sum_j\int_0^\infty [e^{-W(t-s)}]_{ij}\frac{\delta}{\delta z_j^*(s)}\,\dd s.
 \label{eq:qsd}
\end{equation}
The noise has covariance $\E[z_i(t)z_j^*(s)]=K_{ij}(t-s)$ for $t\geq s$ and zero anomalous covariance. For differentiating $D_i$, this formal operator is defined on the full noise history using the exponential continuation. Acting on a causal state, future functional derivatives vanish and the integral reduces to $0\leq s\leq t$. The equal-time source then enters through the commutator below; adding another moving-endpoint term would double count it. This construction extends non-Markovian quantum state diffusion and exponential HOPS~\cite{Diosi1998,Suess2014}.

For $K(t)=e^{-Wt}$ the required identities are
\begin{equation}
 \partial_tD_i=-\sum_jW_{ij}D_j,\qquad [D_i,z_j^*(t)]=K_{ij}(0)=\delta_{ij}.
 \label{eq:D}
\end{equation}
Define unscaled tiers by $\ket{\phi_{\bm k}}=\prod_iD_i^{k_i}\ket{\psi}$, with $\bm e_i$ the unit multi-index. Repeated application of {Equ.~\ref{eq:D}} gives
\begin{align}
 \partial_t\ket{\phi_{\bm k}}={}&\left[-iH_S+\sum_iF_i z_i^*\right]\ket{\phi_{\bm k}}
 -\sum_{ij}k_iW_{ij}\ket{\phi_{\bm k-\bm e_i+\bm e_j}}
 +\sum_i k_iF_i\ket{\phi_{\bm k-\bm e_i}}
 -\sum_iF_i^\dagger\ket{\phi_{\bm k+\bm e_i}}.
 \label{eq:unscaled}
\end{align}
Negative indices contribute zero. The $i=j$ drift is $-k_iW_{ii}\phi_{\bm k}$; $i\ne j$ transfers one hierarchy occupation. Initially $\phi_{\bm0}=\psi_0$ and all higher tiers vanish. Writing $\psi_{\bm k}=\phi_{\bm k}/\sqrt{\prod_i k_i!}$ yields canonical ladder coefficients. In particular, an off-diagonal entry contributes
\begin{equation}
 -W_{ij}\sqrt{k_i(k_j+1)}\,\ket{\psi_{\bm k-\bm e_i+\bm e_j}},\qquad i\ne j.
\end{equation}
Consequently the enlarged generator is
\begin{equation}
 \mathcal G=-iH_S+\sum_iF_i z_i^*+\sum_i(F_i b_i^\dagger-F_i^\dagger b_i)-\sum_{ij}W_{ij}b_i^\dagger b_j.
\end{equation}
The same cross-tier structure appears in generalised HEOM and extended HOPS~\cite{Ikeda2020,Guo2025Extended}. Here $W$ is fixed by a physical mode network and acts alongside distinct local coupling operators. This fixes the index orientation: transposing $W$ can change a complex-hopping model. The hierarchy is exact before truncation and stochastic/time discretisation. Auxiliaries are not independent physical oscillators whose ordinary partial trace gives the system state; the root supplies the physical density.

For the linear hierarchy, the spin density is $\rho_S=\E_{\rm G}[\ket{\psi_{\bm0}}\bra{\psi_{\bm0}}]$, with unnormalised roots sampled under the original Gaussian measure. For a finite ensemble we divide the averaged projector by its pooled trace. Normalising each linear root separately before averaging would change the estimator.

\subsection{Nonlinear importance sampling and normalisation}\label{sec:nonlinear_sampling}
The nonlinear transformation of QSD and HOPS changes the Gaussian sampling measure using the root norm~\cite{Diosi1998,Suess2014,Hartmann2017}. Write the transformed hierarchy as $\ket{\widetilde\Psi}=\sum_{\bm k}\ket{\bm k}\ket{\widetilde\psi_{\bm k}}$. Evaluate the root expectations
\begin{equation}
 \beta_i=\frac{\bra{\widetilde\psi_{\bm0}}F_i^\dagger\ket{\widetilde\psi_{\bm0}}}{\|\widetilde\psi_{\bm0}\|^2}
\end{equation}
and introduce the memory shift
\begin{equation}
 s(t)=\int_0^t e^{-W^*(t-u)}\beta(u)\,\dd u,\qquad \dot s=-W^*s+\beta,\qquad s(0)=0.
\end{equation}
With the same ladder operators and factorial scaling as in Sec.~I\,A, the nonlinear evolution is
\begin{align}
 \partial_t\ket{\widetilde\Psi}&=(\mathcal A-\nu)\ket{\widetilde\Psi},\nonumber\\
 \mathcal A&=-iH_S+\sum_iF_i(z_i^*+s_i)
 +\sum_i\big[F_i b_i^\dagger-(F_i^\dagger-\beta_i)b_i\big]
 -\sum_{ij}W_{ij}b_i^\dagger b_j.
 \label{eq:nonlinear_hierarchy_supp}
\end{align}
The real scalar
\begin{equation}
 \nu=\frac{\mathrm{Re}\bra{\widetilde\psi_{\bm0}}[\mathcal A\widetilde\Psi]_{\bm0}\rangle}{\|\widetilde\psi_{\bm0}\|^2}
\end{equation}
is subtracted from every tier. It keeps the root norm constant in continuous time; small integration drift is removed by a common rescaling. Nonlinear coefficients always use root expectations, even if another code normalises the full hierarchy as an intermediate common gauge.

With the matched nonlinear change of measure, $\rho_S=\E_{\rm nl}[\ket{\widetilde\psi_{\bm0}}\bra{\widetilde\psi_{\bm0}}]$ averages normalised roots. Merely normalising linear trajectories sampled under the original Gaussian measure omits importance sampling and is not equivalent. A full-hierarchy norm is not a physical probability normalisation.

\subsection{Exact stationary grid sampler}
For $W=\Gamma+ih$, $\Gamma\succeq0$, and grid spacing $\delta$, let $R=e^{-W\delta}$. The innovation covariance satisfies
\begin{equation}
 Q=I-RR^\dagger=\int_0^\delta e^{-Wu}(W+W^\dagger)e^{-W^\dagger u}\,\dd u\succeq0.
\end{equation}
Taking $z_0\sim\mathcal{CN}(0,I)$ and independent $\eta_n\sim\mathcal{CN}(0,Q)$ in $z_{n+1}=Rz_n+\eta_n$ gives $\E[z_nz_n^\dagger]=I$ and $\E[z_{n+\ell}z_n^\dagger]=R^\ell$. Zero initial noise would produce a nonstationary covariance.

\begin{quote}\small\ttfamily
Input: Hermitian h, nonnegative amplitude losses gamma, grid spacing delta.\\
W = diag(gamma) + i h; R = expm(-W delta).\\
Q = I - R R-dagger; Hermitise Q and check its eigenvalues.\\
B = eigenvectors(Q) times sqrt(nonnegative eigenvalues(Q)).\\
Draw z = (normal(M) + i normal(M))/sqrt(2).\\
Save z. On each interval draw fresh proper complex xi:\\
\hspace*{1em}z = R z + B xi; save z.\\
Use z-star in HOPS and evolve the matched shift and hierarchy.\\
Estimate observables from the root or derived physical bath projections.
\end{quote}

An eigendecomposition handles semidefinite $Q$; only roundoff-scale negative eigenvalues may be set to zero. Material negative eigenvalues indicate an error, not a request for a positivity repair. A single vector noise process supplies all cross correlations. For general stationary covariance $\Sigma$, replace $Q$ by $\Sigma-R\Sigma R^\dagger$; the continuous diffusion satisfies $BB^\dagger=W\Sigma+\Sigma W^\dagger$.

\subsection{Physical bath observables}
For unshifted factorial auxiliaries, the trajectory estimators are
\begin{equation}
 a_i^{\rm est}=-i\frac{\langle\psi_{\bm0}|\psi_{\bm e_i}\rangle}{\|\psi_{\bm0}\|^2},\qquad
 n_{ij}^{\rm est}=\frac{\langle\psi_{\bm e_i}|\psi_{\bm e_j}\rangle}{\|\psi_{\bm0}\|^2}.
 \label{eq:bath}
\end{equation}
Their averages give $\langle a_i\rangle$ and $\langle a_i^\dagger a_j\rangle$. In the canonical coherent-state representation, a bath annihilation insertion is represented by $-i$ times the corresponding first derivative tier; a normally ordered product uses the bra and ket first tiers. The nonlinear change of measure and common normalisation must be applied to the numerator and root consistently. These are estimator identities after averaging, not an interpretation of the full auxiliary norm as a physical bath density. The $-i$ phase follows the derivative convention in {Equ.~\ref{eq:qsd}} and is checked against explicit modes. All auxiliary coordinates not named are projected to zero. A full-hierarchy number expectation is not this physical projection. These identities apply to passive canonical modes, not arbitrary complex residues without a corresponding mapping.

For different modes, the fourth-order moment used in the main photon figure is
\begin{equation}
 \langle n_i n_j\rangle=\E\!\left[\frac{\|\psi_{\bm e_i+\bm e_j}\|^2}{\|\psi_{\bm0}\|^2}\right],\qquad i\ne j.
\end{equation}
The connected moment subtracts the product of ensemble populations, not the average product of trajectory-level populations. The same ensemble-first prescription is used for connected spin correlations.

\section{Continuous measurement and feedback}\label{sec:feedback}

\subsection{Output measurement and the averaged generator}
Main Fig.~2 uses continuous homodyne monitoring of the middle-cavity output, with feedback applied only to its local spin. We follow the direct-feedback construction~\cite{Wiseman1993} and its conditioned-cavity hierarchy formulation~\cite{Link2022}. Set $\kappa_i=2\gamma_i$, $c=e^{-i\phi}\sqrt{\kappa_2}a_2$ and $G=fS_2^x$. The dimensionless efficiency $\eta_{\rm d}$ is distinct from the anisotropy ratio $g_+/g_-$. With the current convention in the main text, the feedback increment is $\exp(-iG\dd Y)$ and $f$ has units of the square root of frequency. For efficiency $\eta_{\rm d}$, the exact averaged spin--cavity generator has
\begin{align}
 H_{\rm eff}&=H+\frac{\sqrt{\eta_{\rm d}}}{2}(c^\dagger G+Gc),\nonumber\\
 L_{\rm obs}&=\sqrt{\eta_{\rm d}}c-iG,\qquad
 L_{\rm miss}=\sqrt{1-\eta_{\rm d}}c,
 \label{eq:fb_embedding}
\end{align}
and unchanged loss channels on the other cavities. Equivalently, the contribution added to the original Lindblad generator is
\begin{equation}
 \D[G]\varrho-i\sqrt{\eta_{\rm d}}[G,c\varrho+\varrho c^\dagger].
 \label{eq:fb_expanded}
\end{equation}
Both terms follow from applying measurement and feedback in the same stochastic increment. In particular, omitting the quadratic noise drift does not implement the intended controller.

The no-feedback control sets $G=0$. Monitoring without using the output has the same averaged state. The noise-only control retains $\D[G]$ but removes the current-dependent commutator in {Equ.~\ref{eq:fb_expanded}}. It describes a local white-noise Hamiltonian with the same gain, not measurement alone. At fixed nonzero $G$, setting $\eta_{\rm d}=0$ approaches this noise-only control. Gain is not rescaled by $1/\sqrt{\eta_{\rm d}}$.

Connected correlations and concurrence are evaluated on the averaged spin state. For example,
\begin{equation}
G_{ij}^{\alpha\beta}=\E[\langle\sigma_i^\alpha\sigma_j^\beta\rangle_r]
-\E[\langle\sigma_i^\alpha\rangle_r]\E[\langle\sigma_j^\beta\rangle_r].
\end{equation}
The pair moments and single-spin means can be obtained from repeated local spin measurements with the appropriate basis rotations. Averaging a connected correlation or concurrence calculated separately on each trajectory would give a different quantity.

\subsection{Explicit-cavity stochastic Schr\"odinger representation}
Ordinary HOPS noise is not a physical homodyne record. For an independent trajectory check, we use the Bargmann-coefficient representation of the conditioned joint state. For ideal monitoring of every output define
\begin{equation}
 \ket{\chi_{\bm k}}=\frac{i^{|\bm k|}}{\sqrt{\bm k!}}
 \left.\partial_{y^*}^{\bm k}\langle y|\Psi_c\rangle\right|_{y=0},
 \qquad
 \ket{\chi}=\sum_{\bm k}\ket{\bm k}\ket{\chi_{\bm k}}.
\end{equation}
These are phase-rotated physical photon amplitudes, rather than the ordinary coloured-noise auxiliary states. The coupled-mode drift is
\begin{equation}
 \mathcal G_0=-iH_S+\sum_i(F_i b_i^\dagger-F_i^\dagger b_i)
             -\sum_{ij}W_{ij}b_i^\dagger b_j.
\end{equation}
With $C_r=-ie^{-i\phi_r}\sqrt{\kappa_r}b_r$, $\Lambda_r=C_r-iG_r$ and
$\mathcal A=\mathcal G_0-\sum_rG_r^2/2-i\sum_rG_rC_r$, the normalised equation is
\begin{align}
 \dd\ket{\chi}={}&\left[\mathcal A+\sum_r\left(\frac{m_r}{2}\Lambda_r-
                  \frac{m_r^2}{8}\right)\right]\ket{\chi}\dd t\nonumber\\
 &+\sum_r\left(\Lambda_r-\frac{m_r}{2}\right)\ket{\chi}\dd W_r,
 \qquad m_r=\langle C_r+C_r^\dagger\rangle_c.
\end{align}
{For a single controller only the selected $G_r$ is nonzero; distributed control uses several outputs. Full-vector normalisation and the physical trace are}
\begin{equation}
 \sum_{\bm k}\langle\chi_{\bm k}|\chi_{\bm k}\rangle=1,\qquad
 \rho_{S|Y}=\sum_{\bm k}\ket{\chi_{\bm k}}\bra{\chi_{\bm k}}.
\end{equation}
{This all-tier trace applies to the photon-coefficient representation, not to ordinary HOPS auxiliaries. At total degree $|\bm k|\leq K$, it is unitarily equivalent to a total-photon-$K$ wavefunction and provides an independent stochastic benchmark. The reduced-root HOPS construction below instead samples the unconditional spin density.}

The trajectory calculation observes all outputs ideally but uses only the central record for feedback. Averaging the side records gives the same unconditional state as observing only the central output. Its individual conditional states cannot be used to infer entanglement conditioned solely on that central record. The inefficient-detection mechanism comparison uses {Equ.~\ref{eq:fb_embedding}} directly.

\subsection{A reduced-root HOPS representation of the averaged feedback dynamics}
The conditioned coefficient equation above is an explicit-cavity stochastic Schr\"odinger equation (SSE). Its physical estimator traces every photon sector. It is useful as an independent benchmark but is not reduced-root HOPS. For the main feedback figures we instead use the output-field construction below, which samples the \emph{unconditional} spin density through the hierarchy root. A computational trajectory is not identified with an experimental homodyne record. There is no claim here to reconstruct the spin state conditioned only on the measured central output.

The following representation includes every loss port at ideal collection; unactuated ports have $G_r=0$ and are averaged. For inefficient detection, observed and missed ports must be separated. The efficiency mechanism controls in this work use the explicit-cavity master equation instead. For ideal detection, define $\ell_{rj}=e^{-i\phi_r}\sqrt{\kappa_r}\delta_{rj}$. The proper complex cavity and output noises are sampled jointly by
\begin{align}
 \dd z&=-Wz\dd t-\ell^\dagger\dd B, & z(0)&\sim\mathcal{CN}(0,I),\\
 \dd\Xi&=\dd B+\ell z\dd t, & W+W^\dagger&=\ell^\dagger\ell.
 \label{eq:jointoutput}
\end{align}
Here $B$ is proper complex Wiener noise, with $\E[\dd B\dd B^\dagger]=I\dd t$ and $\E[\dd B\dd B^T]=0$, unrelated to the real detector Wiener process in the conditioned equation. For $t>s$ the retarded covariance blocks are
\begin{equation}
 \alpha_{aa}(t,s)=e^{-W(t-s)},\qquad
 \alpha_{oa}(t,s)=\ell e^{-W(t-s)},\qquad
 \alpha_{ao}(t,s)=0.
\end{equation}
Stationarity follows from $-W-W^\dagger+\ell^\dagger\ell=0$. In the retarded output-to-cavity block, the direct input-noise contribution cancels the propagated cavity contribution, giving $\alpha_{ao}=0$. The full output kernel is $\alpha_{oo}(t,s)=\delta(t-s)I$, and the remaining two-time blocks follow by Hermitian symmetry. The output process has marginal white covariance, $\E[\dd\Xi\dd\Xi^\dagger]=I\dd t$, but is correlated with earlier cavity noise. The independent noise-only control deletes both this cross covariance and the corresponding cascade drift; replacing the noise alone would not implement that control. To derive the feedback drift, rewrite one physical loss and its ideal feedback term as
\begin{align}
 \D[c_r]\varrho+\mathcal L_{{\rm fb},r}\varrho
 &=-i[H_{{\rm ser},r},\varrho]+\D[L_r]\varrho,\\
 L_r&=c_r-iG_r,\qquad
 H_{{\rm ser},r}=\frac12(c_r^\dagger G_r+G_rc_r).
\end{align}
Thus the combined channel \emph{replaces} $\D[c_r]$; it is not added on top of that same loss a second time. Its no-jump drift obeys the operator identity
\begin{equation}
 -iH_{{\rm ser},r}-\tfrac12L_r^\dagger L_r
 =-\tfrac12c_r^\dagger c_r-\tfrac12G_r^2-iG_rc_r.
 \label{eq:seriescancellation}
\end{equation}
The $c_r^\dagger G_r$ contributions cancel. Feedback changes the annihilation response without adding an extra cavity-creation source. With the hierarchy's phase convention $a_j\mapsto-i b_j$, the surviving cross term maps to $-\sum_j\ell_{rj}G_rb_j$. The corresponding proper complex output noise uses the same common phase rotation as the ordinary HOPS noise. Together this drift and its causal output covariance reproduce the averaged feedback generator. With the same factorial and phase conventions as the preceding hierarchy, the linear enlarged equation is
\begin{equation}
 \dd\Psi=\left[\mathcal G[z]-\frac12\sum_rG_r^2-\sum_{rj}\ell_{rj}G_rb_j\right]\Psi\dd t
       +\sum_rG_r\Psi\dd\Xi_r^*.
 \label{eq:rootfeedbacklinear}
\end{equation}
Feedback acts on all hierarchy tiers. Only the root enters the reduced-state estimator. Setting every $G_r=0$ recovers the ordinary correlated hierarchy; retaining the white drive and $-G_r^2/2$ while deleting the output--cavity covariance and its drift term gives the independent noise-only control.

The nonlinear equation must change the \emph{joint} Gaussian measure. For coupling operators $A_j$ and vector Gaussian noises $\zeta_j$ with covariance $\alpha$, the norm-squared change of measure shifts
\begin{equation}
 \zeta_i^*(t)\longmapsto\zeta_i^*(t)+
 \sum_j\int_0^t\alpha_{ij}(t,u)^*\langle A_j^\dagger\rangle_u\,\dd u.
 \label{eq:jointGirsanov}
\end{equation}
Apply this once to the combined cavity/output coordinates and couplings $(F,G)$, using the retarded blocks above; the white self-kernel is understood in the prepoint It\^o convention. The cavity block gives the coloured memory shift, the output--cavity block gives its downstream image, and the output delta kernel gives the instantaneous output shift. Let $\beta_j=\langle F_j^\dagger\rangle_0$, $g_r=\langle G_r\rangle_0$ and $\dot s=-W^*s+\beta$. The noise shifts are
\begin{equation}
 z^*\longmapsto z^*+s,\qquad
 \dd\Xi^*\longmapsto\dd\Xi^*+(\ell^*s+g)\dd t.
 \label{eq:feedbackshifts}
\end{equation}
The retarded output shift is fixed by $\alpha_{oa}$; no feedback expectation is added to the forcing of $s$, since the reverse retarded block vanishes. The combined high-tier term $-(F_j^\dagger+\sum_r\ell_{rj}G_r)b_j$ acquires the mean correction $[\beta_j+(\ell^Tg)_j]b_j$. Thus, before a common root normalisation, the nonlinear drift in the same convention is
\begin{align}
 \mathcal A_{\rm fb,nl}={}&\mathcal G[z]+\sum_jF_js_j
 -\frac12\sum_rG_r^2-\sum_{rj}\ell_{rj}G_rb_j\nonumber\\
 &+\sum_j[\beta_j+(\ell^Tg)_j]b_j
 +\sum_rG_r[(\ell^*s)_r+g_r].
 \label{eq:feedbacknl}
\end{align}
The stochastic term remains $\sum_rG_r\Psi\dd\Xi_r^*$. Every tier is divided by the same instantaneous root norm. Because the root norm has a white-noise increment, this normalisation is a stochastic gauge, not simply the deterministic scalar $\nu\dd t$ used for the unmonitored equation. At each step, the discrete implementation evaluates root expectations, advances the memory shift, and evaluates the shifted drift and diffusion using the jointly sampled increments. Couplings to tiers outside the specified total degree are omitted. After the update, all tiers are divided by the same root norm. It must not normalise each tier independently, or normalise linear trajectories without their change of measure.

The prepoint It\^o output shift contains $g\dd t$, not $g\dd t/2$. The equation already contains the no-jump drift $-G^2/2$. Proper complex noise has $\dd\Xi_r^*\dd\Xi_s^*=0$, so importing the usual real-noise Stratonovich correction would change this equation. The finite-step predictor--corrector still requires refinement because its coefficients and coloured fields evolve during the step. The nonlinear HOPS calculations use this uncentred hierarchy.

\section{Controls on the feedback mechanism}\label{sec:mechanism}

The following controls support the physical interpretation of the main figures. They use explicit-cavity master equations or stochastic Schr\"odinger trajectories, as identified in each caption.

\subsection{Feedback history and delayed response}
A local, trace-preserving operation on the intermediate spin leaves the endpoint density unchanged at that instant. Its later effect requires the intervening spin--cavity dynamics. Conversely, stopping feedback leaves the previously prepared joint state intact. The switching protocols in main Fig.~5 test these consequences: delayed activation produces a delayed endpoint response, and endpoint concurrence can continue to rise after feedback stops.

These curves use 4096 linear output-HOPS paths per protocol, with total hierarchy degree $K=5$, $\omega_c\Delta t=0.01$ and noise spacing $0.00125$. Unnormalised root projectors are averaged before dividing by their pooled trace. Pointwise 95\% intervals resample whole blocks, keeping antithetic partners together. Total-photon-cap-five master equations provide the independent reference. Feedback starts at $\omega_ct=10$ or stops at 20 or 30; the delayed-on comparison ends at $\omega_ct=40$ and the switch-off comparisons at 80.

\subsection{Local and nonlocal spin--cavity correlations}
\begin{figure}[!t]
\centering
\includegraphics[width=0.92\textwidth]{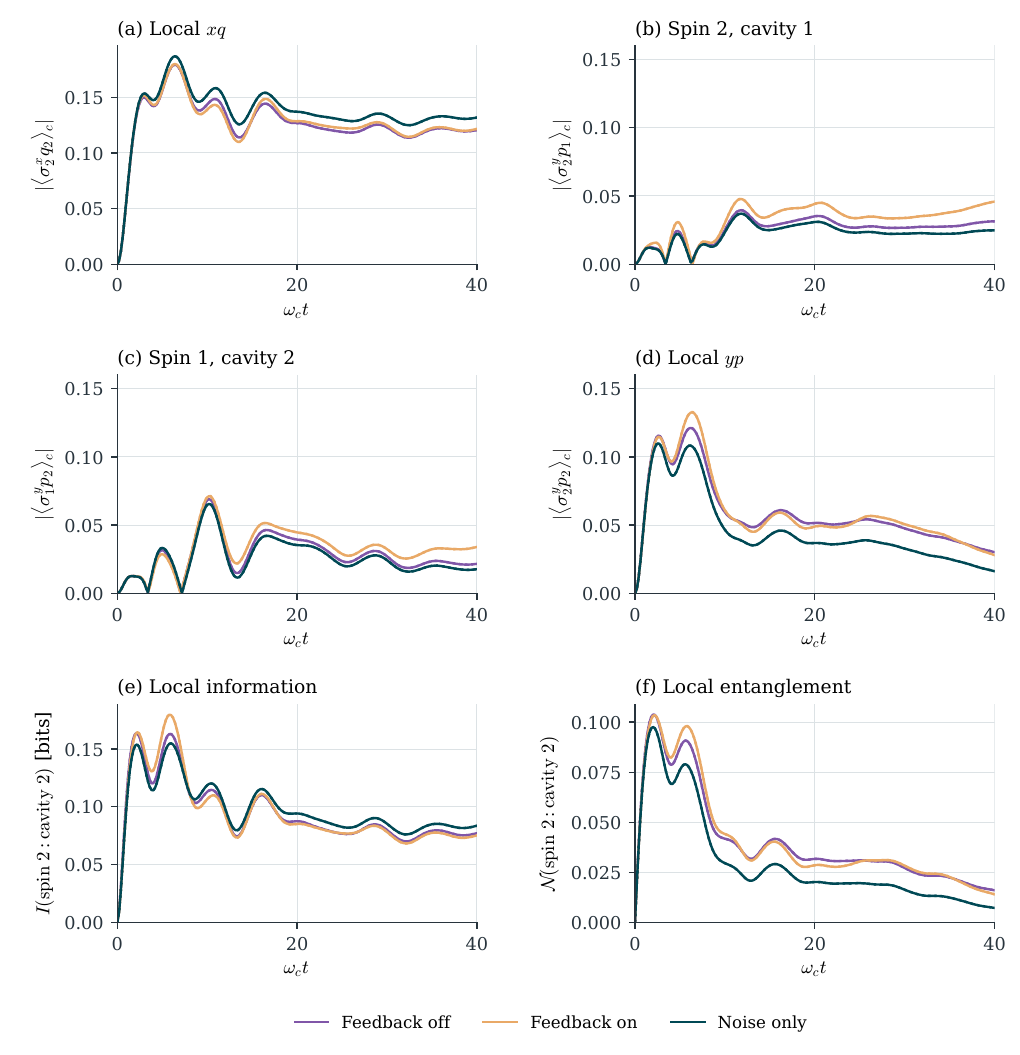}
\caption{Local and remote spin--cavity correlations in the three-spin weak-end chain of main Fig.~2. Each panel compares feedback off, feedback on the middle spin, and independent drive noise on that spin. Panels show (a) $|\langle\sigma_2^xq_2\rangle_c|$, (b) $|\langle\sigma_2^yp_1\rangle_c|$, (c) $|\langle\sigma_1^yp_2\rangle_c|$, (d) $|\langle\sigma_2^yp_2\rangle_c|$, (e) spin-2/cavity-2 mutual information in bits, and (f) their entanglement negativity. Panels (b--d) share a vertical scale. Solid and dotted curves are master equations with total photon caps five and four, respectively; they have no sampling uncertainty.}
\label{fig:spin_cavity_pathway}
\end{figure}

For spin $i$ and cavity $j$, define the mixed covariance $\langle\sigma_i^\alpha X_j\rangle_c=\langle\sigma_i^\alpha X_j\rangle-\langle\sigma_i^\alpha\rangle\langle X_j\rangle$, with normalised quadratures $q_j=(a_j+a_j^\dagger)/\sqrt{2}$ and $p_j=(a_j-a_j^\dagger)/(i\sqrt{2})$. Fig.~\ref{fig:spin_cavity_pathway} shows that feedback changes cross-site spin--cavity correlations without a comparable sustained increase in local middle-spin/cavity correlations, mutual information or negativity. Thus the remote enhancement is not explained by a uniform increase in local spin--cavity entanglement.

\FloatBarrier

The instantaneous feedback source makes this distinction explicit. For a single controller $G=f\sigma_2^x/2$ and $c=-i\sqrt{\kappa_2}a_2$, the additional adjoint generator is
\begin{equation}
 \Delta\mathcal L^\dagger O=\D[G]^\dagger O-i([O,G]c+c^\dagger[O,G]).
\end{equation}
For ideal detection, this gives
\begin{equation}
 \left.\frac{\dd\langle\sigma_2^y p_1\rangle}{\dd t}\right|_{\rm fb}
 =-\frac{f^2}{2}\langle\sigma_2^y p_1\rangle
 -f\sqrt{2\kappa_2}\langle\sigma_2^z p_1p_2\rangle.
 \label{eq:mixed_feedback_source}
\end{equation}
The one-point factors vanish by parity here, so this mixed moment is connected. Its source involves both cavities, not only the reduced spin-2/cavity-2 state. In contrast, the added generator acts trivially on $\sigma_1^y p_2$ and on an endpoint-spin observable: they commute with $G$. Feedback changes their later evolution by changing the joint state on which the other terms act. It is a derivative evaluated on one common state, not the difference between two complete feedback and no-feedback histories.

\subsection{Loss profile and detected output}
\begin{figure}[!t]
\centering
\includegraphics[width=0.92\textwidth]{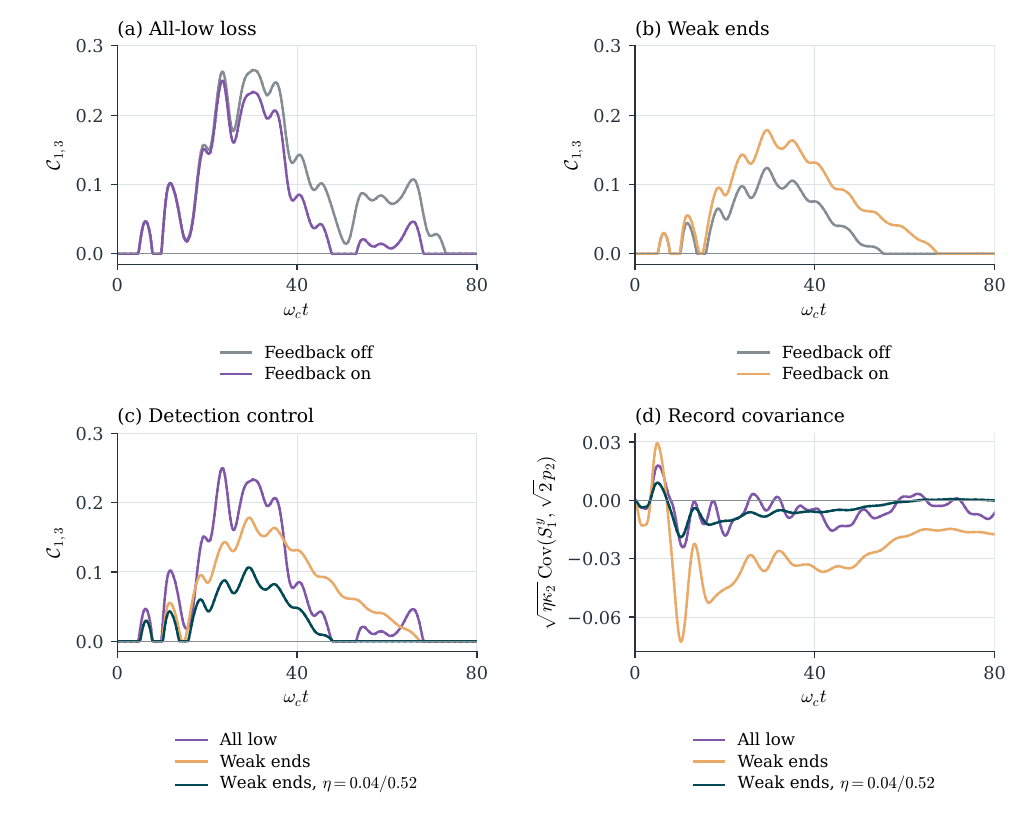}
\caption{Loss and detection controls for the three-spin Rabi chain of main Fig.~2. (a) Endpoint concurrence for uniformly low number losses $(0.04,0.04,0.04)\omega_c$, with feedback off (grey) and on (purple). (b) The corresponding comparison for weak ends, $(0.04,0.52,0.04)\omega_c$, with feedback on in orange. (c) Feedback-on curves from (a,b), together with weak ends at reduced detection efficiency $\eta=0.04/0.52$ (teal). Detection is ideal for the other curves. (d) $\sqrt{\eta\kappa_2}\,\mathrm{Cov}(S_1^y,\sqrt{2}p_2)$ in units of $\sqrt{\omega_c}$, for the three feedback-on cases in (c). Panels (a--c) share a vertical scale. All cases use $J_b=0.25\omega_c$, $f=0.15\sqrt{\omega_c}$ and $\phi=\pi/2$. Solid and dotted curves are master equations with total photon caps six and five.}
\label{fig:loss_efficiency}
\end{figure}

The low-loss control in Fig.~\ref{fig:loss_efficiency} sets every number-loss rate to $0.04\omega_c$, matching the endpoint losses of the weak-end network. Uniformly low loss gives a larger concurrence peak, whereas feedback with a more strongly damped central cavity sustains greater endpoint entanglement over a later interval. The low-loss case subsequently revives, so the comparison concerns this interval rather than a uniform advantage at all times.

Reducing the detection efficiency of the weak-end controller to $\eta=0.04/0.52$ matches the central detected coefficient $\eta\kappa_2$ to the uniformly low-loss case without changing the cavity damping. This reduces the later persistence. Together, these controls support a balance between retaining excitations at the endpoints and coupling the central field to the detector. The spin--detected-quadrature covariance in panel (d) characterises one such correlation; it is not an information rate.

\FloatBarrier

\subsection{Distributed controllers and independent drive noise}
\begin{figure}[!t]
\centering
\includegraphics[width=0.92\textwidth]{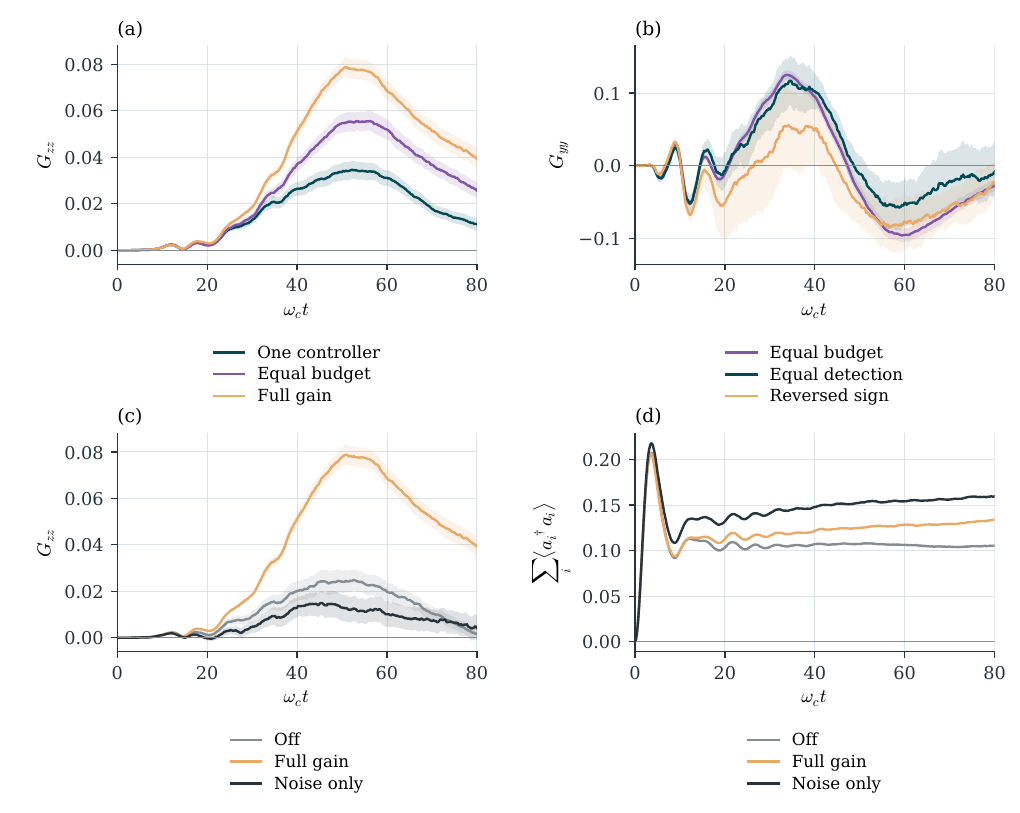}
\caption{Five-spin controller controls from explicit-cavity SSE ensembles, at the weak-end parameters of main Fig.~4. (a) Endpoint $G_{15}^{zz}$ for one controller, three equal-budget controllers, and three full-gain controllers. (b) $G_{15}^{yy}$ for equal budget, equal budget with detection efficiencies $1/3$, and equal budget with the gains on sites 2 and 4 reversed. (c) $G_{15}^{zz}$ with feedback off, full-gain feedback, and independent noise at the same gains. (d) Mean total photon population for the cases in (c). Full gain is orange, equal budget purple and one controller teal; no feedback is grey and independent noise charcoal. In (b), equal detection is teal and reversed sign orange. The off, single, equal-budget, full-gain and noise-only ensembles contain 768, 1536, 768, 1536 and 1024 paths; the two additional controls in (b) each contain 64. Correlation bands are pointwise 95\% whole-block intervals preserving antithetic pairs. Photon moments were stored as batch means, so (d) has no sampling bands. Panels (a,c) share a vertical scale.}
\label{fig:controller_mechanism}
\end{figure}

Fig.~\ref{fig:controller_mechanism} examines how the monitored outputs and local gains affect the five-spin response. The equal-budget control fixes $\sum_j f_j^2$ to the single-controller value while using three outputs. An additional control reduces the three detection efficiencies to $1/3$, matching the summed detected coefficient as well. These controls match coefficients in the evolution equation; they do not equate the information contained in different records.

Reducing detection efficiency or reversing the two outer controller gains changes the endpoint response. Independent drive noise at the full gains can produce more photons than feedback without reproducing its correlations. The remote response therefore depends on the relation between the measured fields and local actuators, not simply on photon population or summed drive strength.

\FloatBarrier

\section{Local bath coordinates and the normal-mode comparison}
\label{sec:normal_modes}
The relevant distinction is between a spatially connected bath with local spin--bath interactions and independent bath coordinates with collective spin couplings. To see this, let $a_i=\sum_\mu U_{i\mu}d_\mu$, where the unitary matrix $U$ diagonalises the Hermitian hopping matrix $h$. The interaction becomes
\begin{equation}
 H_{SB}=\sum_\mu(\widetilde F_\mu d_\mu^\dagger+\widetilde F_\mu^\dagger d_\mu),
 \qquad \widetilde F_\mu=\sum_i U_{i\mu}^*F_i.
\end{equation}
Each $F_i$ acts on one spin, but a normal-mode operator $\widetilde F_\mu$ generally acts on many spins. For uniform damping $\Gamma=\gamma I$, both the mode drift and noise covariance become diagonal:
\begin{equation}
 U^\dagger K(t)U=\diag\big[e^{-(\gamma+i\omega_\mu)t}\big].
\end{equation}
Conventional independent-mode HOPS is then exact for this representation, with independent coloured noises and the collective operators $\widetilde F_\mu$. The alternative used here keeps the physical modes and local operators $F_i$, and represents communication through the off-diagonal bath drift. When losses differ, diagonalising $h$ alone leaves $U^\dagger\Gamma U$ coupled, so the same independent-mode prescription no longer represents the original response.

For $W=\Gamma+ih$ with Hermitian $h$ and $\Gamma$, $[W,W^\dagger]=2i[h,\Gamma]$. Thus $W$ is non-normal when differently damped sites are connected by hopping, and no unitary transformation can diagonalise its full drift. If $U^\dagger hU$ is diagonal, the damping $U^\dagger\Gamma U$ generally is not. The same matrix enters the transformed OU diffusion. A nonunitary similarity transformation that diagonalises $W$ leaves nontrivial noise covariance and noncanonical mode commutators; treating those coordinates as independent canonical damped modes is not an exact basis change. Unequal spin splittings alone do not cause this obstruction, and unequal cavity frequencies with uniform damping still admit an orthogonal normal-mode construction.

The present weak-end case has normal-mode damping, up to eigenvector signs and ordering,
\begin{equation}
 U^\dagger\Gamma U=\omega_c
 \begin{pmatrix}0.14&0&-0.12\\0&0.02&0\\-0.12&0&0.14\end{pmatrix}.
\end{equation}
Its antisymmetric end mode has amplitude lifetime $50/\omega_c$, compared with $10/\omega_c$ for uniform loss. This channel is protected from middle-cavity loss, while the remaining modes and local kernels also change. The longer bare lifetime explains a route to the enhancement, rather than uniquely attributing the interacting dynamics to one mode.

\begin{figure}[!t]
\centering
\includegraphics[width=0.92\textwidth]{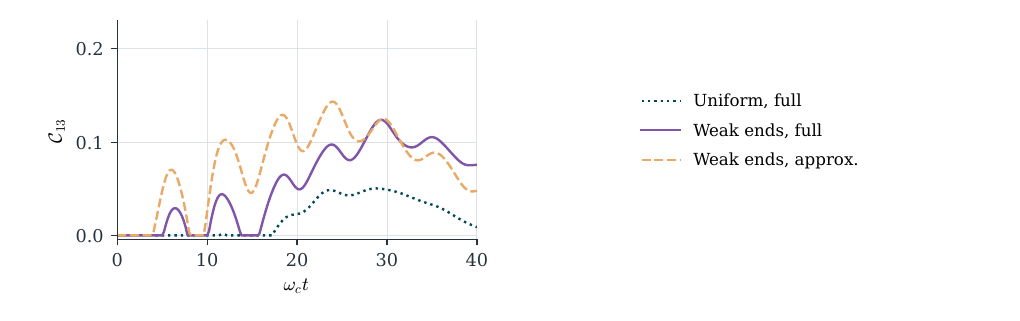}
\caption{Effect of discarding damping couplings between Hamiltonian normal modes. Endpoint concurrence for the full uniform-loss model, full weak-end model, and weak-end independent-normal-mode approximation at the three-spin parameters of main Fig.~1 and $J_b=0.25\omega_c$. All curves use the explicit-cavity reference solver. The approximation retains the exactly transformed spin couplings and diagonal normal-mode damping, but omits its off-diagonal entries.}
\label{fig:normal_mode_control}
\end{figure}

Fig.~\ref{fig:normal_mode_control} compares the full dynamics with an approximation that diagonalises $h$, transforms the spin couplings exactly, but discards the off-diagonal damping in $U^\dagger\Gamma U$. This approximation retains the qualitative weak-end enhancement but changes its magnitude and time dependence. With uniform damping the same normal-mode construction is exact. The distinction is therefore the assumption of independent damped modes, rather than a failure of HOPS to solve the approximate model.

\FloatBarrier

\section{Computational cost relative to explicit-cavity trajectories}\label{sec:cost}
We compare nonlinear HOPS with an explicit-photon stochastic Schr\"odinger equation for seven five-spin regimes used in main Figs.~3 and 4. The Hamiltonians, preparations, loss profiles, feedback protocols and observables are identical between methods. The comparison covers the full evolution to $\omega_cT=80$ and includes both unmonitored and feedback dynamics.

\subsection{Representation size and time evolution}
The dense HOPS and photon bases are truncated by total degree, $\sum_i k_i\leq K_H$ and $\sum_i n_i\leq K_{\rm ph}$, respectively. Each contains
\begin{equation}
 d(N,K)=2^N\binom{N+K}{K}
\end{equation}
complex amplitudes. These are total-occupation bounds, not per-mode cutoffs. For five spins, $K=5$ gives 8064 amplitudes and $K=6$ gives 14784. The selected representations are consequently comparable in size, despite their different physical interpretation and estimators.

Nonlinear HOPS is integrated with a Heun scheme and the normalised photon SSE with Euler--Maruyama. Cutoff and timestep comparisons use nested noise over the full time window. Among the tested timesteps supported by the endpoint correlations and concurrence, HOPS uses $\omega_c\Delta t=0.02$ throughout. The SSE uses $0.01$ for uniform loss with feedback off/on, one weak-end controller and equal-budget controllers, and $0.005$ for weak ends without feedback, full-gain controllers and independent full-gain drive noise. These settings are supported choices, not established global optima or minimal cutoffs.

\begin{figure}[!t]
\centering
\includegraphics[width=0.92\textwidth]{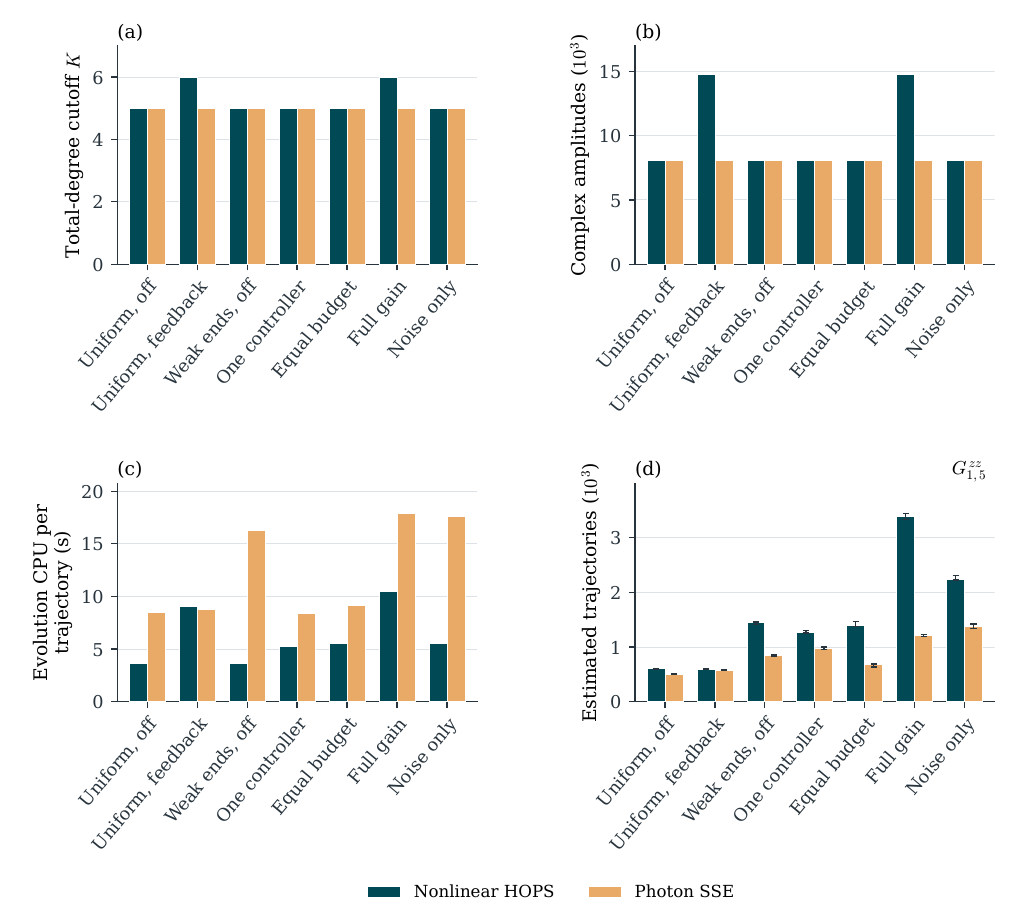}
\caption{Representation and sampling costs for seven five-spin regimes. (a) Selected total hierarchy and photon cutoffs. (b) Corresponding stored complex amplitudes. (c) Measured noise-generation and evolution CPU seconds per trajectory to $\omega_cT=80$. (d) Projected path counts for a pointwise 95\% half-width of 0.005 in $G_{15}^{zz}$ over $20\leq\omega_ct\leq80$. The first two regimes have uniform number losses $0.2\omega_c$; the others use $(0.04,0.92/3,0.92/3,0.92/3,0.04)\omega_c$. A single controller acts on site 3; equal-budget and full-gain controllers act on sites 2--4 with $f_j=0.15\sqrt{\omega_c/3}$ and $0.15\sqrt{\omega_c}$, respectively. Noise only uses independent drives at the full gains. All cases have $J_b=0.25\omega_c$ and otherwise match main Figs.~3 and 4. Sampling projections use large-ensemble variances corrected for the selected timestep. Whiskers in (d) show only uncertainty of this variance correction, not the full uncertainty of the projected count.}
\label{fig:cost_basis}
\end{figure}

Fig.~\ref{fig:cost_basis} compares representation size, per-trajectory CPU time and sampling requirements. Timings are measured on the same macOS ARM64 host with Python 3.10.12, one native thread and batches of 32 paths. They include noise generation and state evolution, but exclude construction, observable recording and file I/O. HOPS generally costs less per path because it permits fewer time steps, although the integrator and work per step also differ. Its larger representation in two feedback cases offsets part of this advantage.

\FloatBarrier

\subsection{Cost at fixed statistical precision}
For each observable, we project the paths required for a pointwise 95\% half-width of 0.005 throughout $20\leq\omega_ct\leq80$. Baseline variances come from the large ensembles used for the physical comparisons; paired samples estimate the change in variance at the selected timesteps. Antithetic partners remain together in the statistical analysis. This criterion concerns sampling precision at individual times, not a simultaneous confidence band or a bound on total numerical error. Concurrence is checked in the numerical comparisons but is excluded from this Gaussian sampling-cost projection.

\begin{figure}[!t]
\centering
\includegraphics[width=0.92\textwidth]{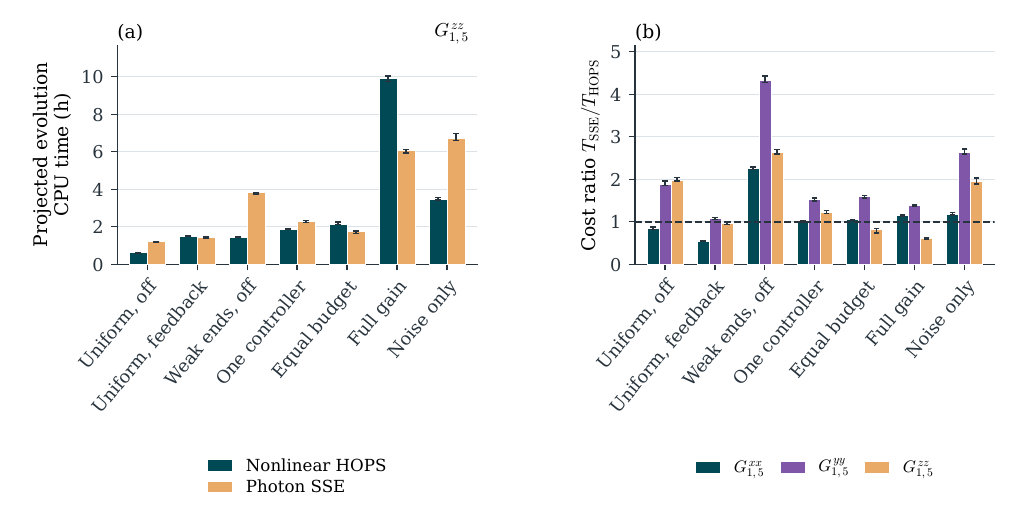}
\caption{Projected CPU cost at fixed statistical precision for the cases in Fig.~\ref{fig:cost_basis}. (a) CPU time per trajectory multiplied by the projected count for $G_{15}^{zz}$. (b) The ratio $T_{\rm SSE}/T_{\rm HOPS}$ for $G_{15}^{xx}$, $G_{15}^{yy}$ and $G_{15}^{zz}$ separately. Ratios above one favour HOPS and ratios below one favour the photon SSE. Both panels use the pointwise precision and time window of Fig.~\ref{fig:cost_basis}. Whiskers propagate only uncertainty in the timestep-variance correction. These are observable-dependent projections from measured trajectory times, not measured runtimes of complete ensembles at the target precision.}
\label{fig:cost_total}
\end{figure}

Multiplying the projected path count by the measured CPU time per path gives Fig.~\ref{fig:cost_total}. HOPS often requires more sampling, so its lower per-trajectory cost does not always produce a lower total cost. For example, the projected $G_{15}^{zz}$ cost favours HOPS for weak ends without feedback, but favours the photon SSE for three full-gain controllers. Uniform-loss feedback is close in cost. The ordering also depends on the spin component, as shown by the separate $xx$, $yy$ and $zz$ ratios. These results establish advantages for both implementations in different regimes, rather than a uniform speedup or basis-size reduction for either method.

\FloatBarrier

\section{Reproducibility}
The numerical data underlying the figures and companion Python and Julia implementations are publicly available at \url{https://github.com/SFlann/correlated-hops}. Additional simulation and analysis code is available from the author upon reasonable request. The records identify physical parameters, numerical settings and ensemble sizes separately for HOPS and independent references. The performance-study settings above are distinguished from those used to generate the main-text ensembles.

\bibliography{references}